\documentclass{aa}
\usepackage{graphicx}       
\usepackage{natbib}

\def\H0{{\rm ~km~s^{-1}~Mpc^{-1}}}

\def\ergsec{{~erg~s$^{-1}$}}

\def\kms{km~s$^{-1}$}
\def\micron{$\mu$m}
\def\whz{{Watt~Hz$^{-1}$}}

\def\hii{\ion{H}{II}}                    
 
\def\ha{{H$\alpha$}}

\def\pa{{Pa$\alpha$}}

\def\p9{{Pa$9$}}

\def\kprime{{$K^\prime$}}
\def\mkprime4{{M$_{\rm K^\prime}$(4kpc)}}
\def\mrprime4{{M$_{\rm R}$(4kpc)}}

\def\L2-10{L$_{\rm 2-10keV}$}

\def\.25{0.25 keV\thinspace}

\def\irf{$IRAS~F$} 
 
\def\d19{D$\,\leq\,$19~Mpc}

\let\lesssim=\la
\let\gtrsim=\ga
 
\let\arcdeg=\degr
 
\newcommand\dummytable{\refstepcounter{table}} %

\begin{document}

\title{The AGN Content of Ultraluminous IR Galaxies:
       High Resolution VLA Imaging of the IRAS 1~Jy ULIRG Sample}
\subtitle{}

\author{
   Neil M. Nagar
   \inst{1,2}
   \and
   Andrew S. Wilson
   \inst{3,4}
   \and 
   Heino Falcke
   \inst{5}
   \and
   Sylvain Veilleux 
   \inst{3,6}
   \and 
   Roberto Maiolino
   \inst{2}
}
\offprints{Neil M. Nagar}
\institute{ 
            Arcetri Observatory, Largo E. Fermi 5,
             Florence 50125, Italy \\
             \email{neil@arcetri.astro.it,maiolino@arcetri.astro.it}
            \and
            Kapteyn Institute, University of Groningen,
             Landleven 12, 9747 AD Groningen, The Netherlands \\
             \email{nagar@astro.rug.nl}
            \and
            Department of Astronomy, University of Maryland,
             College Park, MD 20742, U.S.A. \\
             \email{wilson@astro.umd.edu, veilleux@astro.umd.edu} 
            \and
            Adjunct Astronomer, Space Telescope Science Institute,
             3700 San Martin Drive, Baltimore, MD 21218, U.S.A. \\
            \and
            Max-Planck-Institut f\"{u}r Radioastronomie,
              Auf dem H\"{u}gel 69, 53121 Bonn, Germany \\
             \email{hfalcke@mpifr-bonn.mpg.de}
            \and
	    Current address: 
	    320-47 Downs Lab., Caltech, Pasadena, CA 91125 and
	    Observatories of the Carnegie Institution of Washington, 813 Santa
	    Barbara Street, Pasadena, CA 91101, U.S.A. \\
	     \email{veilleux@ulirg.caltech.edu}
         }
\date{Received April 18, 2003; accepted July 10, 2003}

\abstract{
This paper presents the results of a high resolution radio imaging survey
  of 83 of the 118 ultraluminous infrared galaxies (ULIRGs) in the IRAS 1~Jy
  ULIRG sample. We have observed these ULIRGs at 15~GHz 
  with the Very Large Array (VLA). We find that $\sim$75\% of Seyferts 
  (both type 1 and 2) and LINERs have radio nuclei which are compact at
  our 150~mas resolution. 
  The detection rate of \hii\ nuclei is significantly
  lower (32\%); the detections among these are preferentially 
  \hii\ + LINER/Seyfert composite nuclei.
Among ULIRGs with multiple optical or near-IR nuclei our observations
  detected only one (or no) nucleus; in these the radio detection is typically 
  towards the brightest near-IR nucleus.
The compactness of the radio sources, 
  the higher detection rates in AGN-type nuclei than \hii\ nuclei, 
  the preferential detection of nuclei with unresolved point sources in the
  near-IR, 
  the low soft X-ray to nuclear radio luminosity ratio (arguing against thermal
  emission powering the radio nuclei), and the lack of correlation between radio 
  power and \ha\ luminosity, 
  all support an origin of the detected radio nuclei in AGN related activity.
  This result is especially interesting for LINER ULIRGs
  for which signatures of AGNs have often been ambiguous in other wavebands.
Such a high incidence of AGN would provide, for the first time, a large sample
  in which to study the interplay between AGN, starbursts, and galaxy
  mergers.

\keywords{accretion, accretion disks --- galaxies: active --- galaxies: jets
--- galaxies: nuclei --- radio continuum: galaxies --- surveys}
} 

\titlerunning{AGN radio cores in ULIRGs}
\authorrunning{Nagar, N. M. et al.}

\maketitle

\section{Introduction}

The incidence of active galactic nuclei (AGN) and their relative importance,
as compared to starbursts, in powering the far-IR emission of Ultraluminous Infrared 
Galaxies (ULIRGs; $L_{\rm IR} \geq 10^{12} L_{\sun}$) have been at the center of
a long standing debate. The discovery of probable high-z counterparts of ULIRGs - 
the dusty (presumably star forming) galaxies detected by SCUBA and MAMBO
\citep[e.g.][]{baret00} - gives the debate a new importance. There are now
strong indications that the integrated light from the SCUBA sources and ULIRGs 
accounts for most or all of the sub-millimeter and far-infrared background 
\citep[e.g.][]{cowet02,smaet02}.
Observational evidence that starbursts power ULIRGs includes the following:
starbursts are present in almost every ULIRG, the well known 
correlation between radio and far infrared (FIR) luminosity 
observed in starbursts is followed \citep{yunet01},
and large PAH equivalent widths (empirically related to starburst
activity) are found in most ULIRGs investigated \citep{genet98,lutet98}.
The main evidence for AGNs in ULIRGs as a class comes from optical
spectroscopy:
roughly half of all ULIRGs show nuclear emission lines with ratios 
characteristic of Seyfert or LINER nuclei; of these half of the
Seyfert type nuclei show broad ($\sim5000-10000\,$\kms) permitted lines 
\citep[\ha\ or \pa;][]{veiet99a,veiet99b}. 
ULIRGs with ``warm'' (IRAS 25\micron\ to 60\micron\ flux ratio $>$ 0.2) 
infrared colors generally have Seyfert-like optical
or near-infrared spectra, while hyperluminous 
($L_{\rm IR} \geq 10^{13} L_{\sun}$) objects all have warm infrared colors 
and most have Seyfert 2 spectra and/or ``hidden'' broad line regions
\citep[see e.g.][and the proceedings of the Ringberg conference on ULIRGs, 
1999, \apss, Vol. 266; hereafter 
Ringberg]{veiet95,veiet99a,veiet99b}.
The results listed above, and those from mid-IR spectroscopy \citep{riget99,traet01} 
suggest a dividing line at $L_{\rm IR} \sim 10^{12.5} L_{\sun}$ with ULIRGs above
this energetically dominated by AGN, and those below by starbursts.
Direct observations of the AGN, such as its high brightness temperature radio 
nucleus and jets have been made in very few cases (see below). Hard X-ray
observations, unfortunately available in only a handful of cases, can
directly tell if an AGN \citep[e.g. Mrk~231;][]{galet02}  or starburst
\citep[e.g. NGC~3256;][]{liret02a}, or both 
\citep[e.g. NGC~6240;][]{liret02b,komet03}
power the ULIRG.

ULIRGs are by nature dusty objects and their nuclei are expected to be heavily 
obscured in many cases. Detection of the nucleus itself is, therefore, 
best done at radio or hard X-ray wavebands, which are less affected by obscuration
than the UV to near-IR. An advantage of the radio over the X-ray band lies in
the much higher angular resolution available. A disadvantage of radio in 
comparison with X-ray observations is the negligible luminosity of the former;
for this reason radio observations can never elucidate the dominant power
source (starburst or AGN) in a ULIRG. 
Nevertheless, it is now clear that even low-luminosity accreting black holes 
have detectable compact radio cores \citep[e.g.][]{naget02}. 
These compact flat-spectrum radio cores are usually interpreted as the 
synchrotron self-absorbed base of the jet which fuels larger-scale radio 
emission \citep{blakon79,falbie96}.  
High resolution high frequency radio observations naturally
pick out such AGN-related emission (compact and flat spectrum) while
discriminating against starburst related emission (extended 
and steep-spectrum). Such a survey is therefore the most promising 
way to detect highly obscured accreting black holes embedded in the
starforming environment of ULIRG nuclei. 
There has been no such large survey of ULIRGs so far. A few 
ULIRGs are known powerful radio sources, (e.g. Mrk~231 \citep{ulvet99} and 
3C~273 \citep{manet00}) and the AGNs in these objects have been well studied. 
Another four ULIRGs were studied by \citet{smiet98a} as part of a sample
of Luminous Infrared Galaxies (LIGs; $L_{\rm IR} \geq 10^{11} L_{\sun}$) and 
found to have radio cores with brightness temperature $\gtrsim\,10^6\,$K, which is
not high enough to claim unequivocally the presence of an AGN \citep[see][]{conet91}. 
Of the 43 objects observed by \citet{craet96} in the radio, only four are luminous
enough to fit our definition of a ULIRG; one of these, 03521+0028, is a member of the
sample we study in this paper. A further 2 (of 7) ULIRGs were found to have 2.3~GHz 
cores which are compact at 150~mas resolution above a detection limit of $\sim$5mJy 
in the survey of Seyfert ULIRGs by \citet{royet98}.
Despite their lower IR luminosities, about half of the LIGs studied in the above
papers have a mas-scale radio core, a signature of an AGN.
There is evidence that, with increasing IR luminosity, the AGN may increasingly dominate
the energetics (Ringberg). Thus, one might expect a high incidence of dominant AGNs within 
a sample of ULIRGs.

We have, therefore, started a systematic high resolution radio survey of a well
defined sample of ULIRGs. Here we present results of the first phase of
the project: a 15~GHz VLA survey.
For consistency with \citet[][hereafter VKS]{veiet02} we adopt 
$H_0\,=\,75~$km~s$^{-1}$~Mpc$^{-1}$  and 
$q_0\,=\,0$ 
throughout this paper.

\section{Sample Selection and Observations}

As the parent sample, we use the complete flux-limited IRAS 1~Jy sample \citep{kimsan98} 
which consists of 118 ULIRGs drawn from the IRAS Faint Source Catalog.
The selection criteria of Kim \& Sanders were 
$\delta > -40${\arcdeg},
Galactic latitude $|{b}| >$ 30{\arcdeg},
60~\micron\ flux $f_{60}~> 1$ Jy,
$f_{60} > f_{12}$ (to exclude stars),
warm IR colors (log ($f_{60}/f_{100}) > -0.3$), and
$L_{\rm IR} \geq 10^{12} L_{\sun}$ (i.e. a ULIRG).
The 1~Jy sample is a reliable, complete collection of ULIRGs in the local universe,
but is still small enough to be surveyed in a reasonable amount of time.  
The latest results on optical and IR imaging and spectroscopy of the sample as
a whole can be found in \citet[][hereafter KVS]{kimet02}, VKS and 
references therein. All but one show signs of recent or ongoing 
interactions or mergers (VKS). 

The subsample of the IRAS 1~Jy sample that we have observed in the radio, and report
on here, comprises all 67 ULIRGs with 
$\delta\,>$ 0{\arcdeg}, and 
another 16 ULIRGs with 
$0{\arcdeg}\,>\,\delta\,>\,-20{\arcdeg}$,
as scheduling permitted. 
The median redshift of these 83 ULIRGs is $z_{\rm median}\,=\,0.147$.
In addition we observed two ULIRGs not in the IRAS 1~Jy sample: 
\irf00262+4251 from the `Genzel ULIRG sample' \citep{genet98} and 
\irf05246+0103 which was found to have a gigahertz peaked radio spectrum by 
\citet{craet96}. 

The observations were made at 15~GHz with the Very Large Array 
\citep[VLA;][]{thoet80} in A-configuration during a 
24 hour run on 2002 February 22-23. Weather conditions were good
and the atmospheric phase was stable through the whole run. 
Source positions used for the ULIRGs, taken from the near-IR positions
listed in KVS, were expected to be accurate to $\lesssim$~1{\arcsec}. For each 
ULIRG with multiple nuclei, the phase center was chosen to be the average near-IR 
position of the nuclei and, in all cases, all nuclei fell within the 
effective field of view of the A-array at 15~GHz (including bandwidth 
smearing and time averaging losses). For most
sources, a five minute observation on source was sandwiched between two 1~min
observations on a nearby phase calibrator. Only strong, `good' (`S' or
`P'; see the calibration web page at www.nrao.edu) phase calibrators were used and
the atmospheric variation was slow enough that the phase solutions for the phase
calibrators could be reliably used for the sources in all cases. For the strongest
($>\,0.5\,$Jy) target sources, phase referencing was not used. For optimal $uv$ 
coverage, we obtained two scans on each source separated by at least 1.5~hr for all
but four sources. For 
\irf04313$-$1649, \irf05246+0103, \irf20414$-$1651, and \irf21219$-$1757,
only one scan on source was obtained. 
The sources were observed at elevations of 45{\arcdeg}--80{\arcdeg} except for those
with $\delta\,<\,-10{\arcdeg}$ which had to be observed at elevations as low as 
35{\arcdeg}.

Standard calibration procedures within AIPS were followed, including
antenna gain correction and atmospheric opacity correction.
Variations of antenna gain with elevation were corrected by means of the 
latest antenna gain curves. A sky opacity of 1.5\% was assumed, which
is roughly what is expected at 15~GHz in good weather conditions.
Two four minute observations of 3C~286 (made at elevations of 61{\arcdeg} and 
74{\arcdeg}) were used to set the flux density scale. 
Maps were made with the task IMAGR and, for sources stronger than about 3~mJy,
iterative self-calibration was used to further improve the signal to noise
in the map. The final noise in the maps is about 0.2~mJy and the 
resolution is $\sim$150~mas, equivalent 
to a linear resolution of 50~pc to 800~pc (420~pc at the median redshift)
for the ULIRGs observed here.
Fluxes of detected sources were measured by fitting a single Gaussian using the task 
JMFIT. The highly accurate near-IR positions (KVS) made it easy to identify nuclear 
radio sources even at the $\sim\,3\sigma$ level.

\section{Results}

Our results are summarized in Table~1. 
Column (1) gives the source name,
column (2) the activity type, 
columns (3) and (4) the radio positions of the detected source(s) from our 15~GHz 
   observations, 
column (5) the offset between the 15~GHz radio position and that of the 
   near-infrared nucleus (KVS), 
columns (6) and (7) the peak and total flux densities, 
column (8) the redshift, 
column~(9) the (assumed isotropic) monochromatic radio power and 
column (10) any comments.

\subsection{Detection Statistics of 150~mas Radio Nuclei}

\begin{figure}
\includegraphics{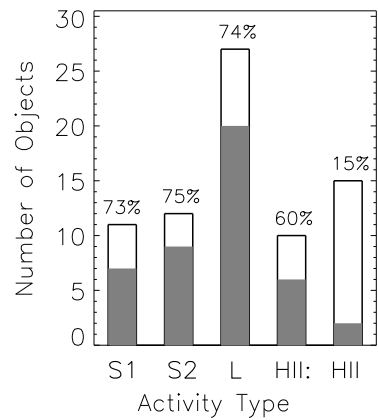}
\caption{Histogram showing the detection rates of 15~GHz radio nuclei for ULIRGs 
with nuclear emission lines characteristic of (from left to right)
Seyfert 1s, 
Seyfert 2s, 
LINERs, 
transitional Seyfert/LINER plus \hii\ regions, and
pure \hii\ regions. 
The total and filled histograms show all radio-observed and all radio-detected 
ULIRGs, respectively. Activity type classifications are from \citet{veiet99a}.
The radio detection rate is significantly higher among Seyferts and LINERs than
among pure \hii\ nuclei.
}
\label{figdetrate}
\end{figure}

\begin{figure*}
\resizebox{\textwidth}{!}{
\includegraphics{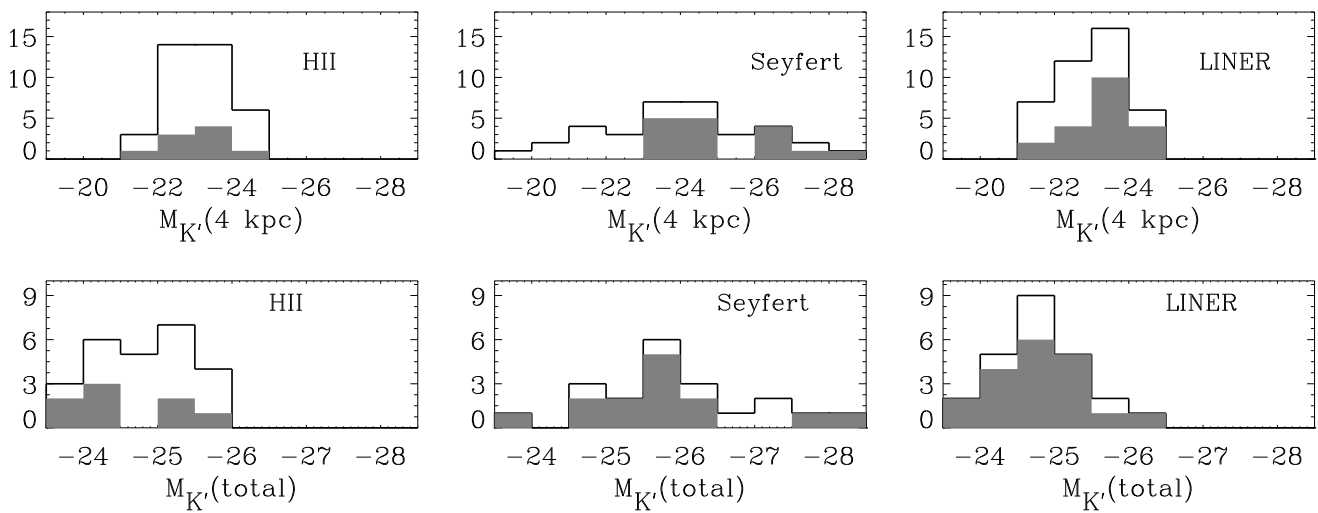}}
\caption{Histograms of 
\textbf{(top)} absolute magnitude of the inner 4~kpc of the ULIRG component in the 
     \kprime\ band (i.e. AGN $+$ galaxy), when available;
\textbf{(bottom)} absolute magnitude of the (total) ULIRG in the \kprime\ band.
Data are from VSK and KVS.
The total and filled histograms represent all ULIRGs (or all ULIRG components)
observed and detected in the radio by us, respectively.
Note that the horizontal scales are different between top and bottom panels.
The radio detection rate increases with increasing nuclear (AGN plus central 4~kpc
of nucleus) \kprime-band luminosity but is independent of the \kprime\ luminosity of the
total ULIRG.
}
\label{figdetstat}
\end{figure*}

Our detection rate is high: radio sources are found towards 47 of 83 ULIRGs 
above a limit of $\sim$0.8~mJy. The nuclear origin of these radio sources 
is supported by the close (within the errors in most cases - Table~1, col. 5) 
positional match to the corresponding near-IR nuclei.
The detection rate is especially high for ULIRGs with AGN-like optical 
spectra: $\simeq$75\% of all Seyferts and LINERs are detected in the radio 
(Fig.~\ref{figdetrate}). 
The detection rate for ULIRGs with 
transitional \hii\ + Seyfert/LINER spectra (see Table~1) is similarly high
(6 of 10 nuclei). 
In contrast, the detection rate of ULIRGs with a pure \hii\ type spectrum is
significantly lower: only 2 of 15 or 14\% (Fig.~\ref{figdetrate} and Table~1).
Additionally, 3 of 8 ULIRGs with unknown 
nuclear spectral classification are detected in the radio. 
There is a large range in redshift among the radio-observed ULIRGs (Table~1);
our non detection of radio cores in the Seyfert nuclei of the luminous
(M$_{\rm K} < -26.5$~mag) galaxies 
\irf11119$+$3257, \irf13218$+$0552, and \irf23498$+$2423 
(see Fig.~\ref{figdetstat}) may be due to them lying at the higher end of the 
redshift range of the sample (Table~1).

Given the above detection rates and the preference of AGNs to reside in `warm' 
ULIRGs (VKS), it is not surprising that the radio detection rate of warm 
ULIRGs (12 of 17) is higher than that of `cool' ULIRGs (35 of 66).
However, with respect to the mid-IR ISO spectral classification 
\citep{lutet99} we detect starburst ULIRGs (7 of 8) at the same rate as 
AGN ULIRGs (7 of 9). Both the above differences in detection rates are not 
reliable due to the small number of objects.
The detectability of a radio nucleus is also related to the 
presence of an unresolved (w.r.t. the point spread
function of the ground based images in KVS and VKS) near-IR nucleus:
20 of 33 (61\%) nuclei with an unresolved near-IR nucleus are
detected in the radio in contrast to only 10 of 53 (19\%) nuclei
without unresolved  near-IR nuclei. Further, the radio detectability is 
related to a higher \kprime\ luminosity of the nucleus in the
central 4~kpc (AGN plus galaxy; \mkprime4) but not to the total
\kprime\ luminosity of the ULIRG (Fig.~\ref{figdetstat}). The former
dependence is significant at the 99.9\% level for all nuclei in the sample or 
for only the Seyfert and LINER nuclei.
No dependence of radio source detectability could be found 
on any of the following: 
interaction class (an estimate of the merger or interaction stage as 
defined in VKS), galaxy morphology, or R or \kprime\ absolute magnitude of 
the total ULIRG. 

\begin{figure}
\includegraphics[width=8cm,angle=-90]{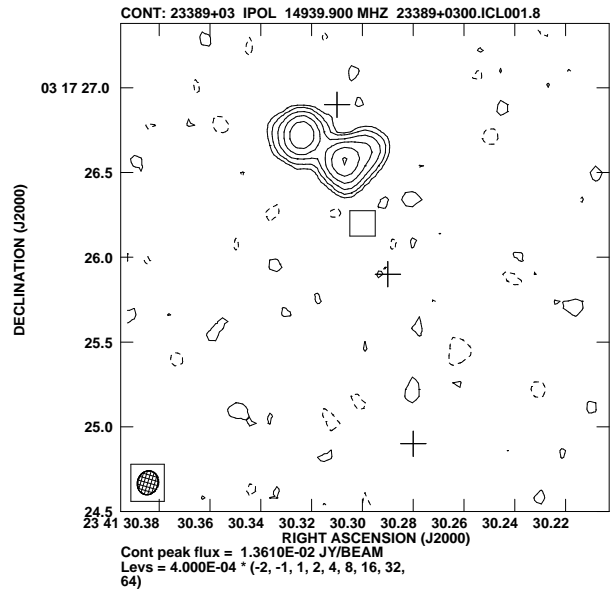}
\caption{15~GHz (2~cm) VLA map of \irf23389+0300 (4C~03.60). The crosses mark the positions
 of the two near-IR nuclei and their average position, and the rectangle marks the position 
 of the northern optical nucleus (KVS).  The southern optical nucleus lies off the bottom 
 of the figure.
}
\label{fig4c03.60}
\end{figure}

Most of the detected radio cores are compact. Our observations are not 
expected to detect extended emission given the limited signal to noise ratio of 
most detections and the `snapshot' nature of the $uv$ coverage.
The two clear exceptions are \irf15327+2340 (Arp~220) which is discussed below
and \irf23389+0300 (4C~03.60) which has twin slightly resolved radio `lobes' 
(Fig.~\ref{fig4c03.60}) separated by 0{\farcs}3 (830~pc at D~=~620~Mpc). 
A comparison with lower resolution fluxes at other frequencies 
\citep[e.g. the 1.4~GHz flux from the FIRST VLA radio survey;][]{whiet97} 
shows that these `lobes' probably have a steep spectrum. The optical and near-IR 
images of 4C~03.60 show it to have two nuclei separated by $\sim$2{\arcsec}, and 
the radio lobes are centered on a point close to the northern near-IR and
optical nucleus (Fig.~\ref{fig4c03.60}).

It is interesting to examine the radio source detections in ULIRGs
with multiple optical or IR components.
Of the radio-detected ULIRGs, VKS (using ground based near-IR images) have 
classified 18 as having two components and 3 as having three components.
We find no dependence of radio source detectability on nuclear separation
between the multiple near-IR nuclei.
In 19 of these 21 ULIRGs, only one of the two or three near-IR 
nuclei is detected in our radio observations. The remaining two
ULIRGs are worth remarking on here.
Arp~220 has two well known \citep{smiet98b} extended radio nuclei with separation 
$\sim$0{\farcs}96, corresponding to the two near-IR nuclei. The exact registration 
between radio and near-IR nuclei is discussed in \citet{scoet98}.
In \irf13539+2920 we have potentially detected radio sources toward both near-IR 
nuclei though the detection of the SE nucleus is not reliable. 
The radio source in most (15 out of 21) of the radio-detected ULIRGs with 
multiple near-IR nuclei coincides with the near-IR nucleus with the brighter
\mkprime4.
In three other cases, the multiple nuclei have similar (to within
$\sim$0.2~mag) \mkprime4.
Only in the three remaining cases are the radio-detected nuclei the fainter 
(w.r.t. the other nuclei in the ULIRG) in \mkprime4:
\irf03521$+$0028 (1~mag fainter in \mkprime4\ though only 0.1~mag fainter in \mrprime4), 
\irf09116$+$0334 (this is an unusual case as the radio source is not towards the bright 
Seyfert galaxy but instead towards the visually fainter but more compact absorption 
line galaxy), and \irf11180$+$1623 (fainter in \mkprime4\ but more compact).

\subsection{Are the Radio Sources AGN Related?}

At the resolution of the observations presented here only three nuclei, all
well known AGNs, have radio brightness temperatures $>$10$^6$~K. 
The other detections have typical brightness temperature limits 
T$_{\rm b}\,>\,10^{2.5-3.5}$~K, much lower than the limit 
\citep[at least 10$^6$~K;][]{conet91} required to confidently invoke
an AGN instead of a highly absorbed nuclear starburst.
Another potential source of the radio emission is radio supernova(e)
\citep[RSN;][]{weiet02} whose high brightness temperature can mimic an
AGN in our observations.
It is therefore relevant to explore whether the compact radio cores we detect 
trace AGN, nuclear starburst activity, or radio supernovae.

A comparison of nuclear radio fluxes with the soft (0.5--2~keV) X-ray emission 
from ULIRGs is a powerful test of whether both originate in thermal 
bremsstrahlung. 
The median nuclear monochromatic luminosity of the radio-detected ULIRGs is 
L$_{\rm 15GHz}$(median)$\,\sim\,10^{30}$ \ergsec~Hz$^{-1}$, so
\citep[following the argument outlined in][]{falet00} thermal bremsstrahlung
would imply 0.5--2~keV luminosities around $\sim\,10^{46}$
\ergsec, significantly higher than observed in ULIRGs \citep{bol99,braet02}.
High column densities could absorb the nuclear soft X-rays but the detection
of broad \ha\ in several of the sample argues against such high columns
at least in the Seyfert~1s.
Also, the high frequency we used is more biased towards detecting flat or 
inverted spectrum sources (typical of AGN nuclei) rather than extended steep
spectrum sources (common among starbursts) as illustrated by our previous
results with low-luminosity AGN \citep{naget02}.
The compactness of the detected radio nuclei ($\leq\,$150~mas or typically 
$\leq\,$50--800~pc for the ULIRGs observed) also argues against extended starbursts.
Another argument for the radio nuclei being AGN related (rather than starburst or
RSN) is the preferential detection of Seyfert and LINER ULIRGs as opposed to 
\hii\ nuclei ULIRGs even though both subgroups have similar median FIR luminosities.
Also, no clear correlation is seen between the nuclear 
($\sim\,2{\arcsec}\,\times\,2{\arcsec}$ aperture) H$\alpha$ luminosity and the radio
power of the nuclei, arguing against both originating in the same source.

It is notable that Arp~220, with its well known starburst dominated 
nuclei, was detected as extended in our survey, and thus the radio nuclei 
\citep[which originate in at least 15 RSN;][]{smiet98b,smiet99} were not 
mistaken as AGN-related. 
The radio-detected ULIRGs in our survey have 
z$_{\rm median}\,= 0.136$ and P$_{\rm median}^{\rm 2cm}\,=$ 22.84 \whz;
powering such a `median' ULIRG would require
$\gtrsim$10 RSN \citep{weiet02} all at peak radio power or 
$\gtrsim$18 of the typical RSN found in Arp~220.

It is worth emphasizing the importance of obtaining deeper radio observations in order 
to detect the other nuclei in ULIRGs with multiple near-IR nuclei. 
VKS have shown that the luminosity ratio of the brightest to faintest
component in multiple component ULIRGs is typically less than 4. 
The preferential radio detection of the more luminous near-IR nuclei thus leads us to
believe that radio observations a factor of a few deeper should 
be able to detect multiple nuclei in several of the ULIRG systems.
Milli-arcsec observations of the VLA-detected ULIRGs can both confirm their AGN 
nature and perhaps find binary massive black hole systems at $<\,$150~mas separations.
Our ongoing followup VLBI observational campaign will attempt to do this.

\section{Conclusions}

We have observed 83 of the 118 ULIRGs in the IRAS 1~Jy sample at 15~GHz with the VLA 
at a resolution of 150~mas and a flux limit of $\sim$0.8~mJy. Our detection rate of
nuclear radio cores is high: about 75\% of all LINER and Seyfert type ULIRGs
are detected. 
The detection rate is significantly lower (only 2 of 15 or 14\%) for
ULIRGs with a pure \hii\ nuclear spectrum.
For multiple component
ULIRGs only one (if any) nucleus - typically the brighter near-IR one - was detected 
in our snapshot observations. Deeper observations to detect binary or multiple 
AGN are highly desirable.
Several factors argue for an origin of the radio emission in AGN
rather than starburst-related activity or radio supernovae. These are:
the compactness and high power of the radio nuclei, 
the preferential detection of AGN-type nuclei than \hii\ nuclei, 
the preferential detection of nuclei with unresolved point sources in the
near-IR, 
the low soft X-ray to nuclear radio luminosity ratio (arguing against a thermal 
origin for the radio emission), and the lack of correlation between radio power 
and \ha\ luminosity.
Such a high incidence of AGN in ULIRGs opens an avenue
to study the inter-relationships between starbursts, AGNs and the galaxy merging
process.

\acknowledgements
This work was partially supported by the Italian Ministry for University and 
Research (MURST) under grant Cofin00-02-36 and the Italian Space Agency (ASI) 
under grant 1/R/27/00. 
This research was supported in part by NASA through grant NAG81755 to the
University of Maryland.
NN thanks the Raman Research Institute for hospitality
during the writing of a part of this paper.
SV thanks the California Institute of Technology and the Observatories of the 
Carnegie Institution for their hospitality.

\clearpage

\begin{table}
    \dummytable 
    \label{tabvla83}
\end{table}

\clearpage

\begin{figure}
\includegraphics[width=20cm,clip]{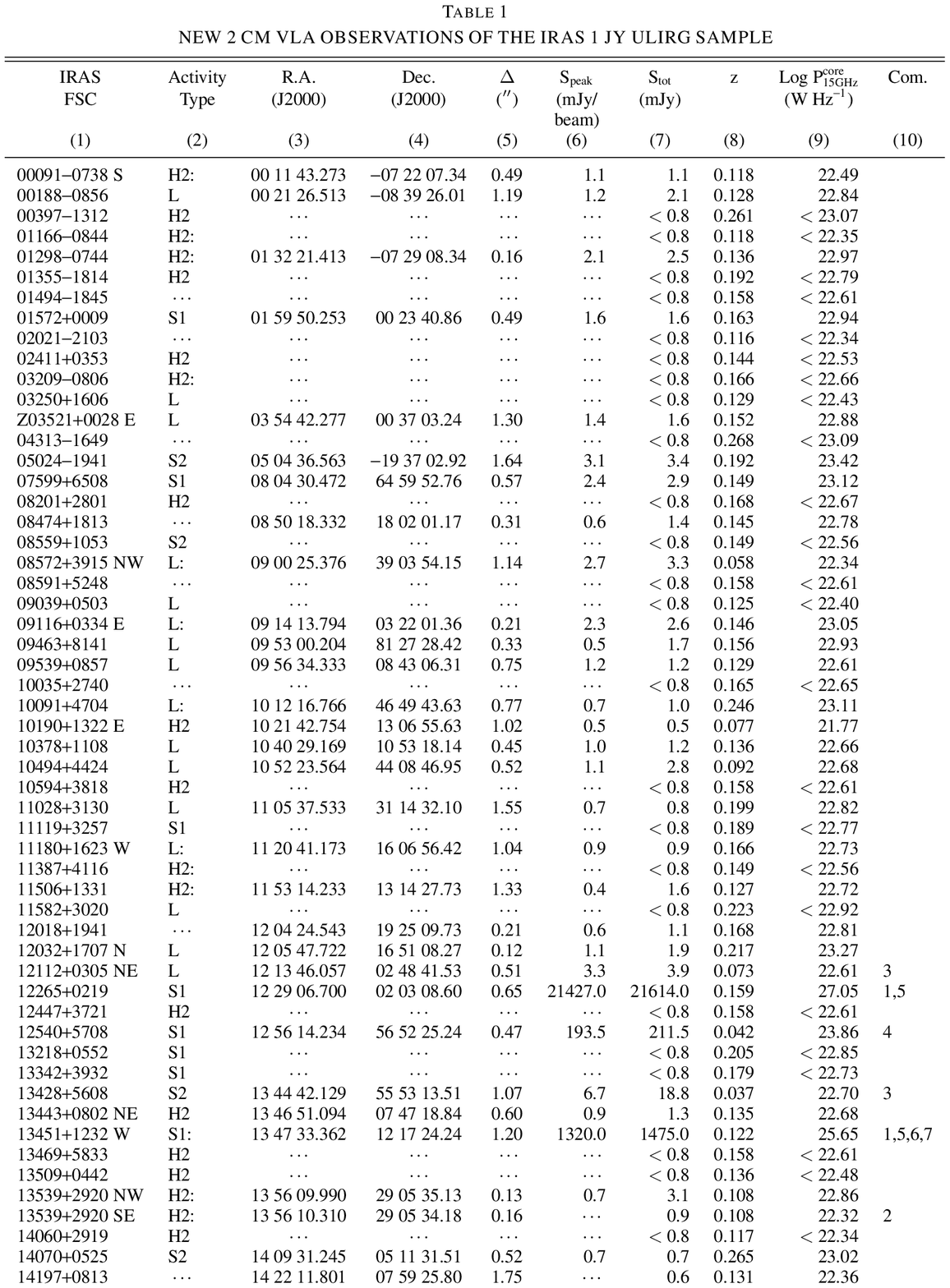}
\end{figure}

\clearpage

\begin{figure}
\includegraphics[width=20cm,clip]{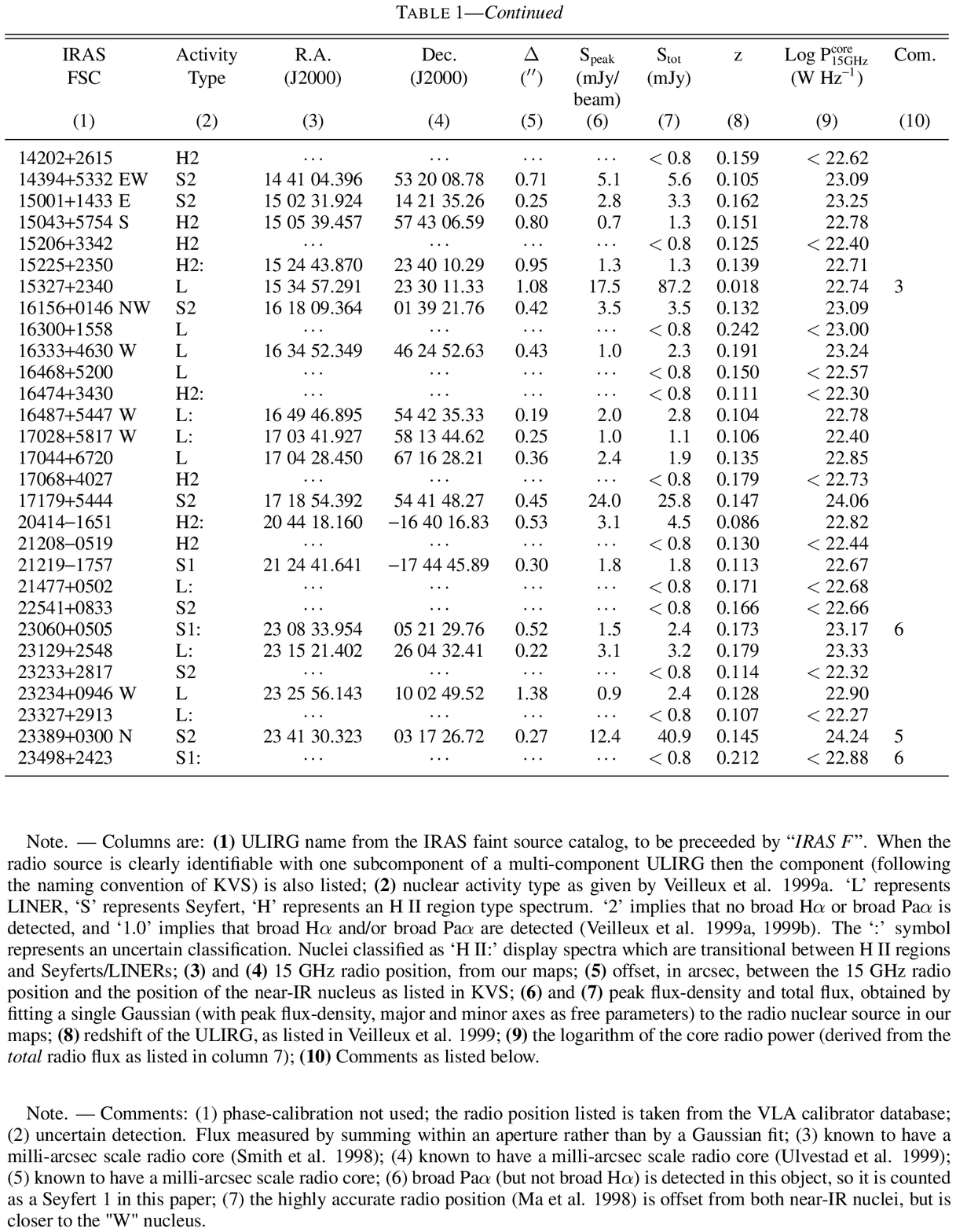}
\end{figure}

\end{document}